\documentclass{article}
\usepackage{iclr2027_conference,times}
\usepackage{graphicx}
\usepackage{fvextra}
\usepackage{tcolorbox}
\usepackage{wrapfig}
\usepackage{xfp}
\tcbuselibrary{breakable,skins}
\tcbset{promptbox/.style={enhanced,breakable,colback=black!2,colframe=black!45,colbacktitle=black!12,coltitle=black,fonttitle=\small\bfseries,boxrule=0.4pt,arc=1mm,left=2mm,right=2mm,top=1.5mm,bottom=1.5mm,before skip=10pt,after skip=10pt}}

\usepackage{amsmath,amssymb}
\usepackage{booktabs,multirow,colortbl}
\makeatletter
\@ifundefined{insert@pcolumn}{\let\insert@pcolumn\insert@column}{}
\makeatother
\usepackage{algorithm,algorithmic}
\usepackage[hyphens]{url}
\usepackage{hyperref}
\hypersetup{hidelinks}
\newcommand{\yangyang}[1]{#1}
\title{MIRA: A Musical Intent Refinement Agent for Aligning Text-to-Music Generation with User Intent}
\author{%
\parbox{0.95\textwidth}{\centering
{\small
Zekai Liu$^{1}$ \quad
Zhilin Wang$^{2}$ \quad
Xuzheng He$^{3}$ \quad
Yu Cheng$^{4}$ \quad
Yang Yang$^{5}$
}\\[6pt]
{\normalfont\small
$^{1}$Shandong University\\
$^{2}$University of Science and Technology of China\\
$^{3}$Central Conservatory of Music \quad
$^{4}$Kunlun Tech Co. Ltd.\\
$^{5}$Shanghai Jiao Tong University
}}}
\iclrfinalcopy
\begin{document}

\maketitle

% Override the template's "Published ... at ICLR 2027" header.
% Must come AFTER \maketitle.
\fancyhead{}
\renewcommand{\headrulewidth}{0pt}

\begin{abstract}

\yangyang{
% Text-to-music systems produce increasingly convincing audio, yet existing evaluation methods reveal little about how well a piece satisfies what the user truly asked for. Global text–audio relevance can overlook implicit musical intent in underspecified prompts and mask failures in specific musical requirements—instrumentation, structure, rhythm, or mood progression. 
% To bridge this gap, we formulate text-to-music intent alignment as satisfying a per-request rubric: a set of independently verifiable items capturing both the request's explicit requirements and its musically implied intent, scored one by one. This enables diagnosis by fine-grained intent source and musical dimension. We instantiate this formulation as MuRA-Bench, a benchmark of user requests sourced from a real-world music generation platform and curated by music experts. 
% We further propose MIRA (Musical Intent Refinement Agent), a test-time agent that optimizes prompts for a black-box text-to-music generator. MIRA grounds a request's implicit intent into verifiable rubrics, then searches over prompt revisions under a bounded generation budget—iteratively generating music, verifying it against the rubrics, and using this feedback to guide a trajectory-aware tree search.
Text-to-music systems produce increasingly convincing audio, yet evaluation reveals little about whether the result matches user intent. A global text–audio relevance score can overlook the implicit intent in underspecified prompts and mask failures in specific requirements—instrumentation, structure, rhythm, or mood progression.
To bridge this gap, we formulate text-to-music intent alignment as satisfying a per-request rubric of independently verifiable items covering both a request's explicit requirements and its implied musical intent. Scoring items individually makes evaluation diagnostic—by intent source and musical dimension—rather than a single opaque score. We instantiate this as MuRA-Bench, a benchmark of real-world platform requests curated by music experts. We further propose MIRA (Musical Intent Refinement Agent), a test-time agent that first grounds a request's intent into rubrics, then searches over prompt revisions for a black-box generator under a bounded budget—iteratively generating music, verifying it against the rubrics, and using this feedback to guide a trajectory-aware tree search.
Experiments across open-source and commercial backends show that MIRA improves intent alignment, enabling an open-source generator to achieve performance comparable to representative commercial 
systems (e.g. Suno and Mureka).
Project page: \url{https://mirareview.github.io/}.}
\end{abstract}

\section{Introduction}

Text-to-music systems increasingly generate high-fidelity, complete songs from natural-language descriptions~\cite{liu2025songgen,lei2024songcreator,yang2025songbloom,lei2026levo2}. Recent systems support lyrics, reference audio, long-form structure, and language-model planning, extending the capabilities of open-source generators~\cite{liu2025songgen,lei2024songcreator,yang2025songbloom,gong2026acestep15,lei2025levo,lei2026levo2}. Yet plausible audio must also realize user intent. Real-world requests combine explicit musical constraints with references to artists, works, scenes, or use cases that convey harder-to-verbalize stylistic choices. Following these requests requires satisfying both stated constraints and contextually implied intent.

Existing evaluations primarily measure audio quality, broad text--audio relevance, or shared musical dimensions, leaving individual request requirements unspecified~\cite{kilgour2019frechet,elizalde2023clap,deshmukh2024pam,huang2025aligning}. Expert-rated datasets, preference platforms, learned quality metrics, and specialized benchmarks strengthen perceptual assessment, but holistic scores do not reveal which request-specific requirements remain unmet~\cite{liu2025musiceval,kim2025musicarena,zhu2026muqeval,ma2026cmirewardbench,wu2026songbench}. A high-scoring track may still omit a required instrument or miss traits implied by a reference. Recent methods decompose audio instructions into verifiable questions or binary rubric items~\cite{kuan2026aqascore,li2026anyaudiojudge}, but primarily address general audio semantics and instruction following rather than expert-revised musical criteria grounded in real-world requests.

We therefore formulate text-to-music intent alignment as \emph{prompt-specific musical rubric satisfaction}. Each request is represented by a set of independently auditable
rubric items covering both explicitly stated constraints and
musical intent inferred from references and context. The specification varies with the request, revealing which requirements are satisfied or missed alongside overall alignment. We instantiate this formulation in \textbf{MuRA-Bench}, a public benchmark containing 100 controlled music-generation requests derived from anonymized intent seeds collected on the Mureka platform. For each request, a language model drafts a musical rubric, which is then reviewed and revised by a music expert. The requests are divided among three experts following shared revision guidelines.Named references are converted into generalized, audibly verifiable, and copyright-conscious musical traits rather than retained as direct imitation targets.

The same rubric perspective also provides an actionable signal for improving generation. Existing methods strengthen prompt following through pre-generation planning or model-specific inference-time control, including reward-guided search, latent optimization, and internal intervention~\cite{gong2026acestep15,novack2024ditto, koo2025smitin,roy2025text2midiinferalign}. We focus on identifying unmet request-specific requirements in completed tracks and converting them into subsequent repairs. We propose \textbf{MIRA}, a \emph{Musical Intent Refinement Agent} that organizes text-to-music generation as verifier-guided search. MIRA uses the request and tool-grounded musical evidence to construct an online rubric, converts generated audio into requirement-level observations with a music-specialized verifier, and explores alternative prompt states under a bounded generation budget. Feedback is retained along the search structure, allowing subsequent actions to preserve satisfied requirements while targeting unresolved ones. Operating outside the generator, MIRA supports structurally different black-box backends without parameter updates.

We validate item-level scores against expert judgments, then compare generators on stated and completed musical intent. We evaluate MIRA on three generation backends under different generation budgets, analyze its major components, and use commercial systems as alignment references. The results show that request-specific evaluation exposes differences obscured by global relevance scores. MIRA consistently improves intent alignment and enables an open-source generator to match or surpass some representative commercial systems under MuRA-Bench evaluation.

Our work makes the following contributions:
\begin{itemize}
\item We formulate text-to-music intent alignment as
\emph{prompt-specific musical rubric satisfaction} and introduce
\textbf{MuRA-Bench}, a public benchmark derived from real-world
platform requests and revised by three music experts. It enables
item-level verification of explicit musical constraints and
expert-validated intent inferred from references and context.

\item We propose \textbf{MIRA}, a musical intent refinement agent that
combines tool-grounded objective construction, requirement-level audio
verification, budgeted prompt search, and feedback memory without
modifying the generation backend.

\item We validate the proposed evaluation through agreement with
expert judgments and demonstrate MIRA across three generation
generators. With commercial systems as references, our results show
that MIRA consistently improves intent alignment and enables an
open-source generator to match some representative
commercial systems.
\end{itemize}

\section{Related Work}

\paragraph{Music Generation Evaluation and Benchmarks.}
Music generation evaluation spans distributional fidelity, text--audio correspondence, perceptual quality, and listener preference~\citep{kilgour2019frechet,elizalde2023clap,deshmukh2024pam,liu2025musiceval,kim2025musicarena,zhu2026muqeval}. FAD characterizes distribution-level fidelity, whereas CLAP measures broad semantic correspondence between text and audio. Expert ratings and preference datasets capture complementary aspects of perceived quality. Reward-model benchmarks also assess musicality and instruction following under text, lyrics, and reference-audio conditions~\citep{ma2026cmirewardbench}. These complementary perspectives characterize system performance; diagnosing intent alignment additionally requires tracing scores to individual musical requirements.

Audio-language models enable more detailed assessment: Audio Flamingo 3 supports audio reasoning, while Music Flamingo specializes in musical properties such as harmony, structure, and timbre~\citep{goel2025audioflamingo3,ghosh2025musicflamingo}. AQAScore measures alignment through targeted audio questions, and AnyAudio-Judge decomposes instructions into binary rubric items~\citep{kuan2026aqascore,li2026anyaudiojudge}. Related text-to-image benchmarks also evaluate verifiable prompt units and examine agreement with human judgments~\citep{hu2023tifa,ghosh2023geneval,kamath2025geneval2}. Our evaluation follows this item-level approach: MuRA-Bench uses expert-revised rubrics derived from real platform requests, separates explicitly stated requirements from intent inferred from references and context, and pairs these criteria with MF-AQA audio-question-answering scores.

\paragraph{Agentic Refinement for Music Generation.}
Text-conditioned audio and music models provide generation backends~\citep{liu2023audioldm,agostinelli2023musiclm,copet2024simple,liu2023audioldm2,liu2025songgen,lei2025levo,lei2026levo2}, with controls over tempo, harmony, instrumentation, vocals, and structure~\citep{melechovsky2024mustango,gong2026acestep15,wang2026segtune}. DITTO further optimizes diffusion noise against differentiable musical objectives without retraining the generator at inference time~\citep{novack2024ditto}. These controls motivate a complementary question: how can feedback on generated audio guide revisions to an open-ended musical request?

Agent research offers mechanisms for such refinement. ReAct supports interaction and external tool use~\citep{yao2023react}; Self-Refine and Reflexion use iterative feedback and experience memory~\citep{madaan2023selfrefine,shinn2023reflexion}; and Tree of Thoughts and LATS search over evaluated reasoning or action trajectories~\citep{yao2023treeofthoughts,zhou2024lats}. GEPA derives reflective updates from execution trajectories, and GEMS combines persistent memory with domain-specific skills in multimodal generation~\citep{agrawal2025gepa,he2026gems}. These methods make feedback, stored experience, and evaluated alternatives available to a planner; music refinement additionally requires observations grounded in the generated audio. MIRA brings tool-grounded intent completion, requirement-level audio feedback, and branch-specific memory into budgeted prompt search, using unmet requirements to guide repairs across music generation backends without updating their parameters.

\begin{figure}[t]
    \centering
    \includegraphics[width=\textwidth]
    {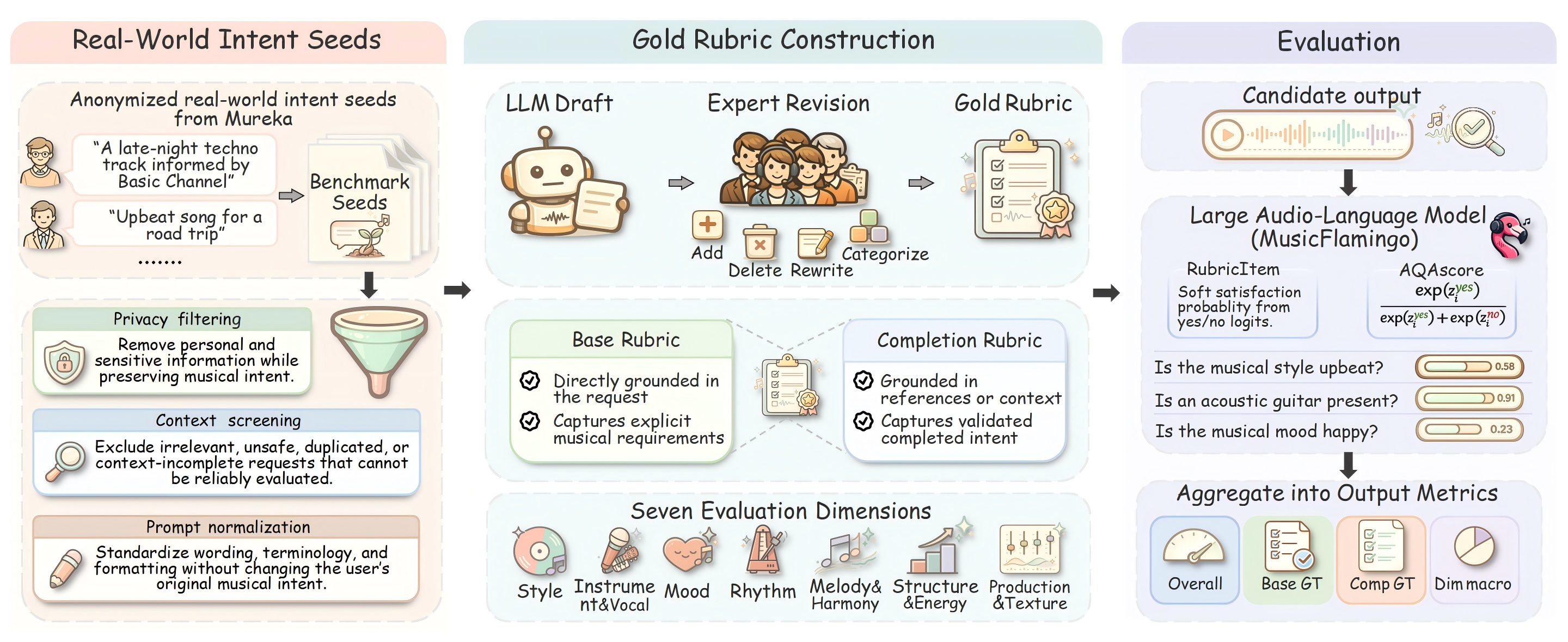}
    \caption{Overview of MuRA-Bench construction and evaluation. Expert-revised, request-specific rubrics support fine-grained assessment with MusicFlamingo. Gold rubrics are used exclusively for evaluation and are not exposed to the generation system.}
    \label{fig:mura-pipeline}
\end{figure}

\section{MuRA-Bench: Rubric-Level Music Intent Alignment}
\label{sec:rubric}

We formulate text-to-music intent alignment as the satisfaction of a
prompt-specific musical rubric. Given a request \(x\), its expert-revised
gold rubric is
\[
R^\star(x)=B^\star(x)\cup C^\star(x),
\]
where \(B^\star(x)\) contains requirements directly supported by the
prompt and \(C^\star(x)\) contains musical traits completed from
references, use cases, and contextual information. Each rubric item
describes one property that can be independently judged from the
generated audio. This formulation accommodates multiple valid musical
realizations without requiring a reference track.

\subsection{Benchmark Construction}

MuRA-Bench contains 100 controlled requests derived from anonymized intent seeds collected on Mureka. Seeds requiring unavailable context, containing sensitive information, or lacking stable musical criteria are excluded. The remaining seeds are normalized while preserving their central intent and realistic expression; raw user logs are not released. The benchmark includes partially specified, reference-based, and hybrid requests with multiple simultaneous constraints.

An instruction-following LLM drafts base and completion requirements and assigns dimensions. Three music experts revise the drafts to preserve explicit constraints, ensure that completed traits are musically plausible and supported by the request or reference context, enforce audible and atomic criteria, and remove or merge redundant or unsupported items. Named artists, songs, and albums are translated into general musical traits rather than direct imitation targets. Items are grouped into seven dimensions: style, instrumentation/vocal, mood, rhythm, harmony/melody, structure/energy, and production/texture. These categories organize request-specific requirements; they do not impose an identical checklist on every request. Figure~\ref{fig:mura-pipeline} summarizes the construction process.

\subsection{Rubric-Based Evaluation}

For a generated track \(a\) and a gold rubric item \(r_i\), we formulate
item-level evaluation as a binary audio question-answering task. The
evaluator is instructed to determine whether the audible content
satisfies \(r_i\) and to answer only \emph{yes} or \emph{no}.
Our final evaluator uses MusicFlamingo, a music-specialized
audio-language model~\cite{ghosh2025musicflamingo}. We compare this
design with direct structured judgment and caption-plus-critic
alternatives through expert agreement experiments.

Following AQAScore~\cite{kuan2026aqascore}, let
\(z_i^{\mathrm{yes}}\) and \(z_i^{\mathrm{no}}\) denote the logits
assigned to the two candidate answers. We define the soft satisfaction
probability as
\[
q_i(a)=
\frac{\exp(z_i^{\mathrm{yes}})}
{\exp(z_i^{\mathrm{yes}})+\exp(z_i^{\mathrm{no}})}.
\]
The probability retains the evaluator's confidence and avoids an
arbitrary threshold when aggregating item-level judgments.

Let \(\mathcal{E}\) denote the set of evaluated request--audio pairs.
For a rubric subset \(S(x)\), let
\(\mathcal E_S=\{(x,a)\in\mathcal E:|S(x)|>0\}\).
We average soft item scores within each audio and then equally across audios:
\[
\operatorname{Score}(S)=
\frac{1}{|\mathcal E_S|}
\sum_{(x,a)\in\mathcal E_S}
\frac{1}{|S(x)|}\sum_{r_i\in S(x)}q_i(a).
\]
Our primary metric is
\(\mathrm{Overall}=\operatorname{Score}(R^\star)\).
We additionally report
\(\mathrm{Base\ GT}=\operatorname{Score}(B^\star)\) and
\(\mathrm{Completion\ GT}=\operatorname{Score}(C^\star)\).
The former measures satisfaction of requirements expressed directly
in the benchmark prompt, whereas the latter measures realization of
expert-revised traits completed from references and context.

To prevent dimensions containing more rubric items from dominating
the analysis, we report \textbf{Dim-Macro}. Each dimension score uses
the same within-audio then across-audio averaging, restricted to audios
with at least one item in that dimension. We equally average the seven scores:
\[
\mathrm{Dim\text{-}Macro}
=\frac{1}{|\mathcal{D}|}
\sum_{d\in\mathcal{D}}
\operatorname{Score}(R_d^\star),
\]
where \(R_d^\star(x)\) contains the gold items assigned to dimension \(d\).
Overall averages all items within each audio, rather than averaging
dimension scores or the Base and Completion scores equally.

For grader validation, music experts rate item-level satisfaction.
We normalize 20-point ratings and exclude uncertain
responses; Appendix~\ref{app:human-assessment} describes the calculation.
These ratings are used only to measure grader agreement. The expert gold rubrics and human grading labels are never
provided to the generation or refinement system.

\section{MIRA: Musical Intent Refinement Agent}
\label{sec:method}

\begin{figure}[t]
    \centering
    \includegraphics[width=\textwidth]
    {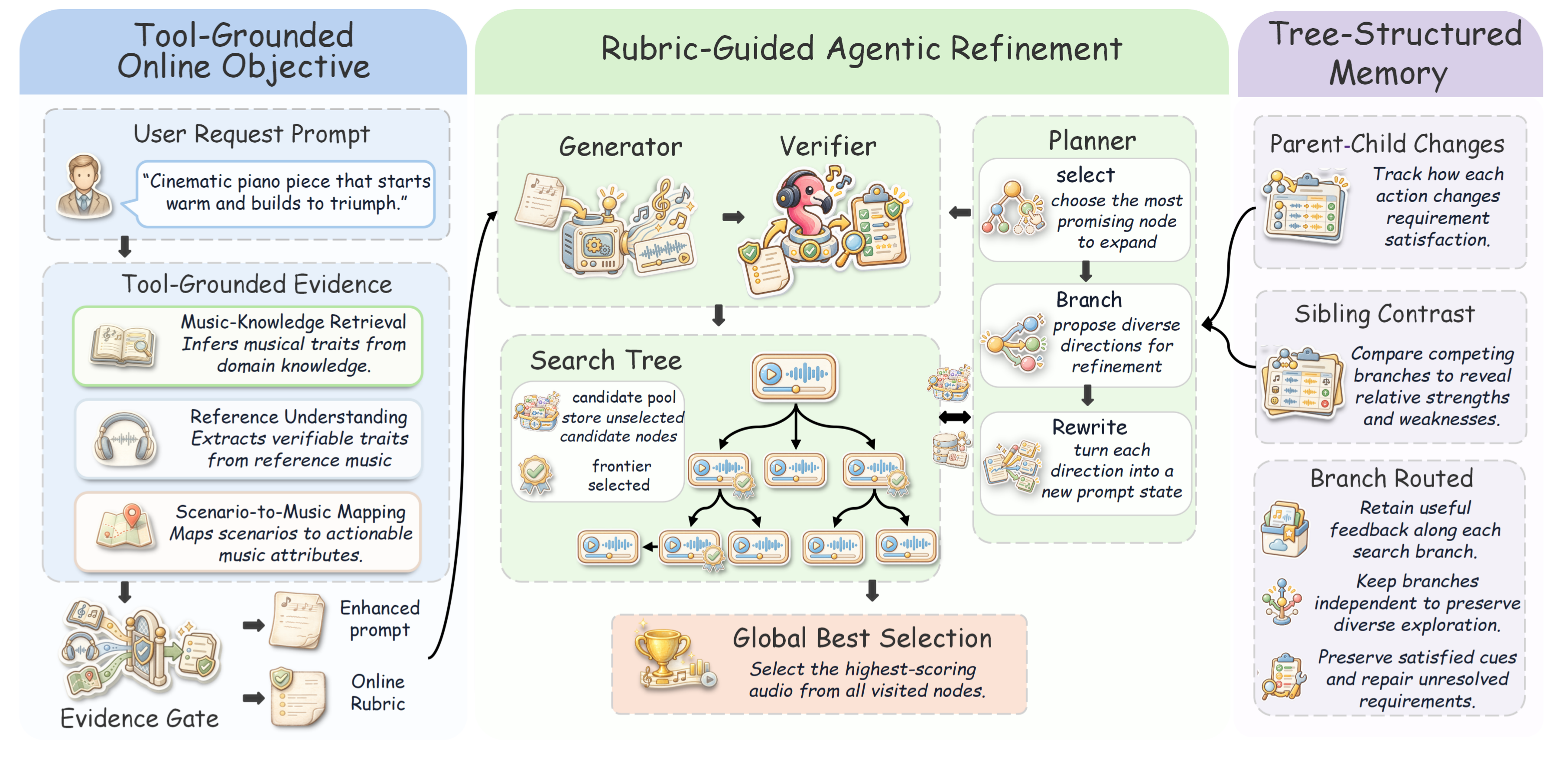}
    \caption{Overview of MIRA and its verifier-guided refinement process.}
    \label{fig:mira-overview}
\end{figure}

\yangyang{We introduce MIRA (Musical Intent Refinement Agent), a verifier-guided agentic framework that collaborates with any black-box text-to-music generator to improve intent-alignment. As shown in Figure~\ref{fig:mira-overview}, 
MIRA is built on three core principles: 1) externalizing latent user intent to grounded, objective, and verifiable rubrics; 2) using tree-search to balance exploration and exploitation; 3) 
maintaining branch-specific memory to preserve rich and the most relevant optimization history.}

\yangyang{
Starting from a user prompt $x$ and a black-box text-to-music generator $G$, MIRA first uncovers the implicit intents in the prompt and translates it into a set of grounded rubrics $R$, together with the musical control parameters $\kappa$ required by the backend generator. Then, MIRA performs a verifier-guided search process to iteratively improve the generation prompt. Specifically, it constructs a search tree
\(\mathcal T=(\mathcal V,\mathcal E_{\mathcal T})\), where each node
\(v\in\mathcal V\) stores
\[
n_v=(p_v,a_v,o_v,M_v),
\]
with \(p_v\) denoting complete generation prompt, $a_v\sim G(\cdot\mid p_v, \kappa)$ the generated audio sample, \(o_v\) the rubric-level judgement produced by a music-specialized verifier, and
\(M_v\) the branch-specific memory accumulated along the path to \(v\).
}

\subsection{Tool-Grounded Online Rubric}

A user request may contain both directly stated constraints and musical
intent that requires external knowledge to interpret. References, usage
scenarios, and functional descriptions can imply musical properties
that are not fully expressed in the prompt. MIRA preserves the original
request as an immutable intent anchor and, when necessary, invokes
external tools for music-knowledge retrieval, reference-audio
understanding, or scenario-to-music attribute mapping.

Tool outputs first pass through an evidence gate. Only information with
a traceable source, supporting evidence, and sufficient confidence is
admitted for intent completion. Named references are converted into generalizable, audibly verifiable musical properties.

Given the request \(x\) and admitted external evidence \(E\), a text LLM
constructs the online musical rubric
\[
R(x,E)=\{r_1,\ldots,r_m\},
\]
where each rubric item \(r_i\) represents an atomic musical requirement
that can be assessed independently from the generated audio. The rubric
is fixed before search and shared by all candidate nodes. It is derived
only from the request and admitted evidence, without access to the
expert gold rubric used for offline evaluation.

\subsection{Rubric-Guided Search}

Text-to-music generation is stochastic, and the same alignment failure
may admit several plausible prompt revisions. A single verification
result therefore does not uniquely determine the appropriate repair.
Moreover, following a single refinement trajectory may improve one
rubric item while degrading others that were previously satisfied.
MIRA therefore formulates prompt refinement as rubric-guided search
under a finite generation budget.

For a node \(v\), MusicFlamingo evaluates every online rubric item and
returns an item-level observation together with an aggregate score:
\[
\begin{aligned}
q_{v,i}
&=P_{\mathrm{MF}}(\mathrm{yes}\mid a_v,r_i),\\
o_v
&=(q_{v,1},\ldots,q_{v,m}),\\
s_v
&=\frac{1}{m}\sum_{i=1}^{m}q_{v,i}.
\end{aligned}
\]
The aggregate score supports candidate comparison, while the item-level
scores identify satisfied and unresolved parts of the online rubric.
Rubric items with \(q_{v,i}<\eta\), where \(\eta=0.60\), are treated as
low-scoring items and may receive concise targeted feedback.

At each search layer, the agent uses each retained parent's prompt,
item-level observation, and branch memory to propose several complete
candidate prompts. Candidates expanded from the same frontier are generated and verified independently; outcomes are shared only after all candidates at that layer have been evaluated.

After evaluating a layer, MIRA retains a bounded search frontier. It
preserves the candidate with the highest aggregate score and a
complementary candidate with stronger mean satisfaction over its
lowest-scoring quarter of rubric items; any remaining positions are
filled according to aggregate score. The observed rubric state thus
determines which candidates are retained and which items are targeted
by subsequent repairs, while the search topology remains controlled by
the generation budget.

Let \(\mathcal V_B\) denote all nodes generated and evaluated under
budget \(B\), including the initial generation. The budget constraint is
\[
|\mathcal V_B|\leq B.
\]
We evaluate a compact setting with \(B=3\) and a broader setting with
\(B=9\). Both use the same search mechanism and differ only in the
available generation budget.

\subsection{Tree-Structured Feedback Memory}

Retaining only the highest-scoring candidate does not explain why a
revision succeeded or whether it improved some rubric items at the
expense of others. MIRA therefore aligns its feedback memory with the
search tree, allowing subsequent decisions to use both the local effect
of a revision and its performance relative to competing search
directions.

For a non-root node \(v\) with parent \(p(v)\), MIRA first measures the
item-level change from its parent:
\[
\Delta^{\mathrm{par}}_{v,i}
=
q_{v,i}-q_{p(v),i}.
\]
It then compares the node with the other candidates evaluated at the
same search layer:
\[
\Delta^{\mathrm{cmp}}_{v,i}
=
q_{v,i}
-
\frac{1}{|\mathcal N_l(v)|}
\sum_{u\in\mathcal N_l(v)}q_{u,i},
\]
where \(\mathcal N_l(v)\) denotes the other candidates visible at layer
\(l\). The contrastive term is omitted when this set is empty. The
parent--child difference captures the local effect of a revision, while
the candidate contrast identifies the relative strengths and weaknesses
of its search direction.

To reduce sensitivity to minor verifier fluctuations, changes whose
absolute magnitude does not exceed \(\epsilon=0.03\) are treated as
stable; larger positive and negative changes are recorded as improvements
and regressions, respectively. MIRA summarizes these signals as
branch-specific memory \(M_v\), recording which rubric items should be
preserved, which remain unresolved, and which revision directions have
produced regressions. The text LLM uses this memory when proposing
subsequent candidate prompts.

Memory is propagated only along the corresponding branch. Different
branches do not access one another's complete prompts or private
trajectories, preserving diversity across horizontal search while
allowing feedback to persist within each branch. All memory is reset
between user requests.

\subsection{Candidate Selection}

Within budget \(B\), MIRA returns the highest-scoring successfully
visited candidate:
\[
v^\star=\arg\max_{v\in\mathcal V_B}s_v,
\qquad
a^\star=a_{v^\star}.
\]
 Aggregate
satisfaction is the primary selection criterion. Ties are resolved
using the rubric-item pass rate, lower-tail satisfaction, the minimum
item score, and finally generation order.

\section{Experiments}
\label{sec:experiments}

\paragraph{Implementation Details.}
We evaluate MIRA on ACE-Step v1.5 Turbo~\citep{gong2026acestep15}, SongGeneration2 Large~\citep{lei2026levo2}, and YuE2-3B~\citep{yuan2025yue,map2026yue2}. Kimi 2.6~\citep{moonshot2026kimik26} serves as the text planner, and MusicFlamingo provides requirement-level verification~\citep{ghosh2025musicflamingo}. For each backend, generation and verification run on a single NVIDIA A800. We compare against ten commercial models using direct prompting through their official APIs: MiniMax music-2.6/3.0~\citep{minimax2026music26,minimax2026music30}, Mureka V9/V9.5~\citep{mureka2026models,mureka2026v95}, StepAudio 3 Music~\citep{stepaudio2026music}, and Suno v4/v4.5/v5/v5.5/v6~\citep{suno2024v4,suno2025v45,suno2025v5,suno2026v55,suno2026v6}. MIRA uses generation budgets of $B=3$ and $B=9$, including the initial generation. Main results cover all 100 MuRA-Bench requests; equal-budget comparisons and component analyses use a fixed subset of 20 requests. Evaluator validation covers 100 clips from 25 requests,while system-level human evaluation covers 400 clips from all 100 requests.

\paragraph{Agreement with Expert Judgments.}
We assess evaluator agreement using 1,536 rubric-item ratings
from 100 clips covering 25 requests.
Three music experts independently annotate assigned subsets,
with system identities and automatic scores hidden;
Appendix~\ref{app:human-assessment} details the protocol.
We compare eight evaluation methods.
Five use AQA scoring based on normalized yes/no logits~\citep{kuan2026aqascore}:
MusicFlamingo (MF-AQA)~\citep{ghosh2025musicflamingo}, Qwen3-Omni-30B-A3B~\citep{xu2025qwen3omni},
Qwen2.5-Omni-7B~\citep{xu2025qwen25omni}, MOSS-Music-8B~\citep{mossmusic2026}, and
AnyAudio-Judge-7B~\citep{li2026anyaudiojudge}.
The remaining methods are MF-Prompting, which directly
elicits five-point requirement ratings from MusicFlamingo;
MF-Cascade, which passes requirement-independent
MusicFlamingo captions to a text judge; and CLAPScore~\citep{elizalde2023clap,wu2023laionclap},
which measures audio--text embedding similarity.

Table~\ref{tab:evaluator-agreement} shows that MF-AQA
achieves the highest agreement across all six metrics.
These results support the use of MusicFlamingo with AQA
scoring for requirement-level evaluation and candidate
selection in MIRA.

\begin{table}[t]
\centering
\caption{Agreement with expert ratings on 100 clips from
25 requests. The first five evaluators use AQA.
MF denotes MusicFlamingo.}
\label{tab:evaluator-agreement}
\begingroup
\fontsize{7.2}{8.4}\selectfont
\setlength{\tabcolsep}{3pt}
\renewcommand{\arraystretch}{1.03}
\begin{tabular}{@{}lrrrrrr@{}}
\toprule
& \multicolumn{2}{c}{Item-level}
& \multicolumn{3}{c}{Clip-level}
& Pair \\
\cmidrule(lr){2-3}\cmidrule(lr){4-6}
Evaluator
& SRCC & $\tau_b$
& LCC & SRCC & $\tau_b$
& Acc.\ (\%) \\
\midrule
MusicFlamingo
& \textbf{.690} & \textbf{.532}
& \textbf{.817} & \textbf{.823} & \textbf{.646}
& \textbf{79.59} \\
Qwen3-Omni-30B-A3B
& .509 & .375 & .632 & .631 & .480 & 74.15 \\
Qwen2.5-Omni-7B
& .469 & .344 & .599 & .643 & .465 & 72.79 \\
MOSS-Music-8B
& .502 & .371 & .591 & .593 & .429 & 62.59 \\
AnyAudio-Judge-7B
& .462 & .335 & .561 & .556 & .393 & 68.03 \\
\midrule
MF-Prompting
& .552 & .460 & .732 & .743 & .567 & 72.79 \\
MF-Cascade
& .436 & .350 & .563 & .608 & .434 & 57.82 \\
CLAPScore
& .245 & .175 & .361 & .331 & .223 & 62.59 \\
\bottomrule
\end{tabular}
\endgroup
\end{table}

\begin{table}[t]
\centering
\caption{Results on the full MuRA-Bench benchmark.
All scores are multiplied by 100; higher is better.
Superscripts in MIRA rows indicate absolute changes from the
same backend's direct-prompt baseline on this scale.
Bold marks the best score within each refinement backend
or among the commercial models.}
\label{tab:main-results}

\begingroup
\definecolor{miraBand}{RGB}{233,242,248}
\definecolor{miraGain}{RGB}{55,125,70}
\definecolor{miraDrop}{RGB}{175,65,65}

\newcommand{\mup}[2]{%
  #1\textsuperscript{\textcolor{miraGain}{\ensuremath{\uparrow}#2}}}
\newcommand{\mdown}[2]{%
  #1\textsuperscript{\textcolor{miraDrop}{\ensuremath{\downarrow}#2}}}

\setlength{\tabcolsep}{2.5pt}
\renewcommand{\arraystretch}{1.18}

\resizebox{\textwidth}{!}{%
\begin{tabular}{@{}l*{11}{c}@{}}
\toprule
& \multicolumn{4}{c}{\textbf{Intent alignment}}
& \multicolumn{7}{c}{\textbf{Musical dimensions}} \\
\cmidrule(lr){2-5}
\cmidrule(lr){6-12}
\textbf{Backend / Setting}
& Overall & D-Macro & Base & Comp.
& Style & I/V & Mood & Rhythm & H/M & S/E & P/T \\
\midrule

\rowcolor{miraBand}
\multicolumn{12}{c}{\textit{ACE-Step v1.5 Turbo}} \\

Direct
& 67.6 & 67.2 & 69.6 & 66.1
& 68.1 & 62.4 & 71.9 & 71.0 & 64.1 & 69.4 & 63.3 \\

MIRA ($B=3$)
& \mup{74.2}{6.6}
& \mup{74.1}{6.9}
& \mup{77.1}{7.5}
& \mup{71.1}{5.0}
& \mup{74.6}{6.5}
& \mup{\textbf{69.5}}{7.1}
& \mup{81.2}{9.3}
& \mup{75.4}{4.4}
& \mup{71.4}{7.3}
& \mup{72.8}{3.4}
& \mup{73.7}{10.4} \\

MIRA ($B=9$)
& \mup{\textbf{76.3}}{8.7}
& \mup{\textbf{75.6}}{8.4}
& \mup{\textbf{79.1}}{9.5}
& \mup{\textbf{73.4}}{7.3}
& \mup{\textbf{75.4}}{7.3}
& \mup{69.1}{6.7}
& \mup{\textbf{83.0}}{11.1}
& \mup{\textbf{77.2}}{6.2}
& \mup{\textbf{72.8}}{8.7}
& \mup{\textbf{74.6}}{5.2}
& \mup{\textbf{77.0}}{13.7} \\

\rowcolor{miraBand}
\multicolumn{12}{c}{\textit{SongGeneration2 Large}} \\

Direct
& 60.5 & 59.2 & 62.9 & 58.3
& 55.4 & 51.1 & 65.0 & 68.9 & 49.0 & 64.1 & 60.7 \\

MIRA ($B=3$)
& \mup{66.2}{5.7}
& \mup{64.9}{5.7}
& \mup{68.9}{6.0}
& \mup{63.2}{4.9}
& \mup{62.0}{6.6}
& \mup{55.8}{4.7}
& \mup{75.0}{10.0}
& \mup{70.2}{1.3}
& \mup{56.2}{7.2}
& \mup{67.7}{3.6}
& \mup{67.0}{6.3} \\

MIRA ($B=9$)
& \mup{\textbf{72.0}}{11.5}
& \mup{\textbf{70.8}}{11.6}
& \mup{\textbf{75.3}}{12.4}
& \mup{\textbf{68.4}}{10.1}
& \mup{\textbf{75.1}}{19.7}
& \mup{\textbf{63.5}}{12.4}
& \mup{\textbf{77.2}}{12.2}
& \mup{\textbf{74.6}}{5.7}
& \mup{\textbf{59.8}}{10.8}
& \mup{\textbf{72.1}}{8.0}
& \mup{\textbf{73.4}}{12.7} \\

\rowcolor{miraBand}
\multicolumn{12}{c}{\textit{YuE2-3B}} \\

Direct
& 76.5 & 75.8 & 79.2 & 73.3
& 77.3 & 70.5 & 83.7 & 74.8 & 70.5 & 77.3 & 76.7 \\

MIRA ($B=3$)
& \mup{80.5}{4.0}
& \mup{79.9}{4.1}
& \mup{83.1}{3.9}
& \mup{77.6}{4.3}
& \mup{\textbf{82.9}}{5.6}
& \mup{77.1}{6.6}
& \mup{86.5}{2.8}
& \mup{79.3}{4.5}
& \mup{73.8}{3.3}
& \mup{79.6}{2.3}
& \mup{80.4}{3.7} \\

MIRA ($B=9$)
& \mup{\textbf{82.1}}{5.6}
& \mup{\textbf{81.3}}{5.5}
& \mup{\textbf{85.0}}{5.8}
& \mup{\textbf{78.4}}{5.1}
& \mup{82.8}{5.5}
& \mup{\textbf{77.7}}{7.2}
& \mup{\textbf{87.7}}{4.0}
& \mup{\textbf{81.9}}{7.1}
& \mup{\textbf{75.8}}{5.3}
& \mup{\textbf{81.0}}{3.7}
& \mup{\textbf{81.9}}{5.2} \\

\rowcolor{miraBand}
\multicolumn{12}{c}{\textit{Commercial systems (direct prompting)}} \\

MiniMax music-2.6
& 75.9 & 75.4 & 78.2 & 74.0
& 76.0 & 68.9 & 81.6 & 79.1 & 67.5 & 78.3 & 76.4 \\

MiniMax music-3.0
& 77.0 & 75.6 & 79.8 & 74.8
& 75.9 & 72.3 & 83.1 & 77.1 & 67.4 & 75.1 & 78.6 \\

Mureka V9
& 80.6 & 80.0 & 84.0 & 77.2
& 83.8 & 75.5 & 86.7 & 81.8 & 73.1 & 80.5 & 78.4 \\

Mureka V9.5
& 81.2 & 80.6 & 83.8 & \textbf{78.9}
& 83.0 & 75.2 & 87.6 & 82.4 & 75.1 & 81.0 & 80.2 \\

StepAudio 3 Music
& 79.6 & 79.3 & 82.6 & 76.5
& 82.2 & 75.0 & 84.9 & 79.6 & 75.6 & 79.5 & 78.2 \\

\addlinespace[2pt]

Suno v4
& 73.6 & 72.8 & 75.8 & 72.9
& 74.5 & 61.7 & 80.7 & 76.2 & 64.7 & 77.3 & 74.7 \\

Suno v4.5
& 76.8 & 76.0 & 80.0 & 73.2
& 78.2 & 69.4 & 84.3 & 76.8 & 65.9 & 80.2 & 77.0 \\

Suno v5
& 79.7 & 79.0 & 83.1 & 76.0
& 81.7 & 72.5 & 85.3 & 81.0 & 73.5 & 80.6 & 78.6 \\

Suno v5.5
& \textbf{81.9} & 80.9 & 85.3 & 77.7
& 84.8 & 75.3 & 86.9 & \textbf{83.3}
& 70.9 & 84.6 & \textbf{80.4} \\

Suno v6
& 81.8 & \textbf{81.5} & \textbf{85.8} & 77.6
& \textbf{85.1} & \textbf{75.6} & \textbf{87.8} & 80.5
& \textbf{76.3} & \textbf{85.2} & 80.2 \\

\bottomrule
\end{tabular}%
}

\par\vspace{3pt}
\raggedright\fontsize{7}{8.5}\selectfont
D-Macro: dimension-macro; Base/Comp.: explicit/inferred intent.
I/V: instrumentation/vocal; H/M: harmony/melody;
S/E: structure/energy; P/T: production/texture.
Direct denotes direct prompting.
\textcolor{miraGain}{$\uparrow$}: increase;
\textcolor{miraDrop}{$\downarrow$}: decrease relative to Direct.
\par
\endgroup
\end{table}

\paragraph{Main Results.}
Table~\ref{tab:main-results} shows consistent alignment improvements across all three generation backends. With $B=9$, MIRA increases Overall by \textbf{12.9\%} on ACE-Step, \textbf{19.0\%} on SongGeneration2, and \textbf{7.3\%} on YuE2-3B relative to direct generation. Overall and D-Macro improve at both budgets, and increasing the budget from $B=3$ to $B=9$ further improves both explicit and inferred intent across all three backends.

Among commercial systems, recent Suno and Mureka models achieve the strongest overall alignment. Version updates show a clear overall trend toward better intent alignment: Suno's Overall score rises from 73.6 in v4 to 81.9 in v5.5 and 81.8 in v6, while MiniMax and Mureka also improve across their evaluated versions. This trend highlights intent alignment as an important axis of progress in music generation. MuRA-Bench captures these advances in models' ability to satisfy request-specific musical requirements.

YuE2-3B with MIRA achieves the highest Overall score among the evaluated systems, reaching 82.1 compared with 81.9 for Suno v5.5, the strongest commercial baseline on this metric. ACE-Step with MIRA also improves from 67.6 to 76.3, compared with 75.9 for MiniMax music-2.6. These results demonstrate that MIRA enables open-source generators to achieve intent-alignment performance comparable to representative commercial systems.

\begin{figure}[t]
    \centering
    \includegraphics[width=\linewidth]{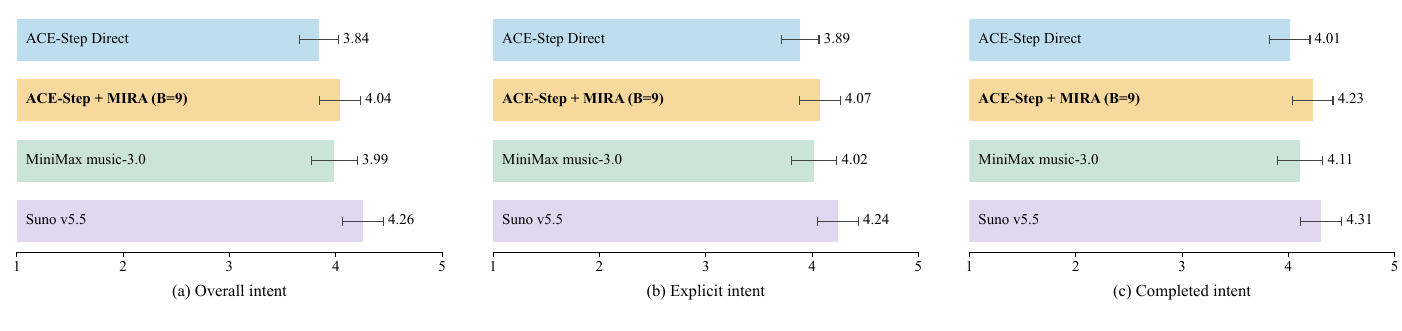}
    \caption{Blinded expert ratings of overall, explicit, and inferred intent on MuRA-Bench. Bars show mean ratings on a 1–5 scale, with 95\% confidence intervals.}
    \label{fig:human-evaluation}
\end{figure}

\paragraph{Human Evaluation.}
Figure~\ref{fig:human-evaluation} complements evaluator agreement with blinded system-level assessment. ACE-Step with MIRA receives higher mean ratings than direct ACE-Step for overall, explicit, and inferred intent. Its means also exceed MiniMax music-3.0 while remaining below Suno v5.5 across all three dimensions. These descriptive results provide complementary human evidence for the alignment improvements observed in automatic evaluation.

\begin{table}[t]
\centering
\caption{Component analysis on a fixed subset of 20 MuRA-Bench
requests.
Direct generation uses $B=1$; all remaining configurations use $B=9$.
Tool grounding is first combined with best-of-nine sampling,
followed by adaptive search without memory and then
tree-structured memory.
Bold marks the best score within each backend.}
\label{tab:component-ablation}
\begingroup
\fontsize{7.2}{9}\selectfont
\setlength{\tabcolsep}{2.5pt}
\renewcommand{\arraystretch}{1.18}

\begin{tabular*}{\textwidth}{@{\extracolsep{\fill}}lc*{8}{c}@{}}
\toprule
& & \multicolumn{4}{c}{\textbf{ACE-Step v1.5 Turbo}}
& \multicolumn{4}{c}{\textbf{YuE2-3B}} \\
\cmidrule(lr){3-6}
\cmidrule(lr){7-10}
\textbf{Setting} & $B$
& Overall & D-Macro & Base & Comp.
& Overall & D-Macro & Base & Comp. \\
\midrule

Direct & 1
& 0.759 & 0.756 & 0.803 & 0.710
& 0.816 & 0.811 & 0.868 & 0.752 \\

Best-of-9 & 9
& 0.800 & 0.795 & 0.874 & 0.719
& 0.848 & 0.850 & 0.910 & 0.775 \\

Tools + Best-of-9 & 9
& 0.815 & 0.807 & 0.871 & 0.748
& 0.860 & 0.852 & 0.911 & 0.800 \\

Tools + Adaptive search & 9
& 0.826 & 0.814 & 0.862 & 0.781
& 0.867 & 0.863 & \textbf{0.913} & 0.808 \\

Full MIRA & 9
& \textbf{0.836} & \textbf{0.826}
& \textbf{0.880} & \textbf{0.782}
& \textbf{0.872} & \textbf{0.868}
& 0.911 & \textbf{0.821} \\

\bottomrule
\end{tabular*}
\endgroup
\end{table}

\begin{wrapfigure}{r}{0.48\textwidth}
    \centering
    \includegraphics[width=\linewidth]
    {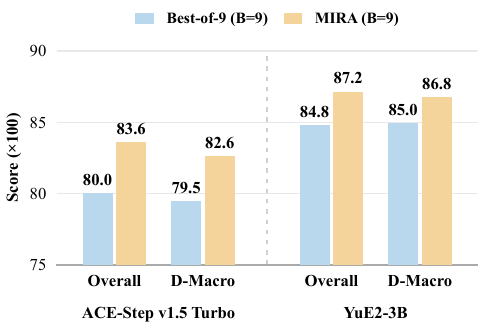}
    \caption{Best-of-9 versus MIRA with $B=9$.
    Scores are multiplied by 100;
    the vertical axis starts at 75.}
    \label{fig:bestof9-vs-mira}
\end{wrapfigure}

\paragraph{Equal-Budget Comparison.}
Figure~\ref{fig:bestof9-vs-mira} compares MIRA with Best-of-9 on the 20-request subset. Both use nine generations and MusicFlamingo for final selection. MIRA improves Overall by 3.6 points on ACE-Step and approximately 2.35 points on YuE2-3B, with higher D-Macro scores on both backends. These results support more effective use of the generation budget than independent sampling.

\paragraph{Component Analysis.}
Table~\ref{tab:component-ablation} examines the contributions of tool grounding, adaptive search, and structured memory on the same 20 requests. Adding tool grounding to Best-of-9 raises Comp.\ from 0.719 to 0.748 on ACE-Step and from 0.775 to 0.800 on YuE2-3B. By grounding intent completion in external musical knowledge, this component provides more informed task specifications that help generators realize musical traits left implicit in the original request.

Adaptive search then improves Overall, D-Macro, and Comp.\ on both backends. ACE-Step's Comp.\ rises from 0.748 to 0.781, illustrating how requirement-level observations guide search toward unresolved aspects of intent more effectively than independent sampling from a fixed specification.

Adding structured memory further improves Overall and D-Macro on both backends, with full MIRA achieving the highest scores on these metrics. These results support using branch-specific records of successful and unsuccessful revisions to guide subsequent search.

\section{Conclusion}

We presented a unified framework for evaluating and improving intent alignment in text-to-music generation. MuRA-Bench converts real-world platform requests into expert-revised rubrics covering explicit constraints and inferred intent, enabling requirement-level evaluation. Agreement with expert judgments supports requirement-targeted audio question answering for fine-grained intent assessment.
MIRA integrates tool-grounded intent completion, verifier-guided search, and tree-structured memory to refine black-box music generation. Experiments across three open-source backends demonstrate consistent alignment improvements, enabling an open-source generator to achieve performance comparable to representative commercial systems. Gains over repeated sampling under equal generation budgets further support the effectiveness of this search framework.
% ICLR 2027 requires an AI use statement. Authors must supply an accurate
% statement covering their research and writing workflow before submission.
\bibliographystyle{iclr2027_conference}
\bibliography{references}

\clearpage
\appendix
\appendix

\section*{AI Use Statement}

We used ChatGPT for grammar checking and text refinement,
and AI coding assistants for portions of the code.
We also used LLMs to generate candidate musical rubrics
during benchmark construction, which were subsequently
revised by music experts, as described in
Appendix~\ref{app:benchmark-details}.
The authors reviewed the AI-assisted materials and
take responsibility for the final content.

\section{MuRA-Bench Construction and Expert Revision}
\label{app:benchmark-details}

\subsection{Benchmark composition}

MuRA-Bench contains 100 controlled requests derived from anonymized
Mureka intent seeds. The benchmark preserves musical intent while
excluding requests that require unavailable context or do not admit
stable evaluation criteria. It contains 68 instrumental and 32 vocal
requests, with 1,566 rubric items: 996 base items and 570 completion
items. Requests contain 6--31 items (mean 15.66; median 14.5).
Table~\ref{tab:app-benchmark-composition} reports the number of rubric
items in each musical dimension. A request can contain items from
several dimensions.

\begin{table}[htbp]
\centering
\small
\caption{Distribution of the 1,566 rubric items across seven
musical dimensions.}
\label{tab:app-benchmark-composition}
\begin{tabular}{lr}
\toprule
Rubric dimension & Items \\
\midrule
Style & 135 \\
Instrumentation/vocal & 303 \\
Mood & 218 \\
Rhythm & 257 \\
Harmony/melody & 114 \\
Structure/energy & 199 \\
Production/texture & 340 \\
\midrule
Total & 1,566 \\
\bottomrule
\end{tabular}
\end{table}

\subsection{Expert revision procedure}

Three music experts revised disjoint subsets of the benchmark
independently. Each expert reviewed the original request, base and
completion requirements, and dimension assignments using a shared
set of revision guidelines. The interface supported editing
requirements and recording revision decisions and comments.

\paragraph{Revision checklist.}
The expert-facing guidelines organize review into five steps:

\begin{enumerate}
    \item \textbf{Check explicit requirements.}
    Compare each base item with the original request. Restore omitted
    constraints, including negative constraints; remove unsupported
    additions and duplicates; and correct truncation or changes in
    meaning.

    \item \textbf{Identify what needs interpretation.}
    Check whether a named artist, work, scenario, use case, image,
    or story actually occurs in the request and needs elaboration.
    If the request is already explicit and complete, leave the
    completion target and completion requirements empty.

    \item \textbf{Check contextual support.}
    Retain completion items that are relevant to the user's goal,
    supported by the identified context, compatible with explicit
    requirements, and assessable by listening. Remove personal
    preferences and overly specific guesses. For reference-based
    requests, retain only relevant traits of that reference; for
    functional requests, prioritize mood, energy, and development
    rather than imposing unnecessary instruments or exact tempi.
    The interface asks reviewers to write one assessable requirement
    per line.

    \item \textbf{Assign dimensions.}
    Organize the revised requirements into musical dimensions without
    inventing additional requirements merely to populate a category.

    \item \textbf{Record the decision.}
    Indicate whether the revised sample is usable and whether
    completion is appropriate or unnecessary. Record reasons for
    weak support, required rewriting, or a recommendation against use.
\end{enumerate}

\subsection{A complete rubric example}

Table~\ref{tab:app-rubric-example} presents a complete benchmark rubric
for battle music in a plains-based role-playing game. Six base items
capture the requested musical elements and atmosphere; six completion
items specify musical traits associated with the gameplay context.

\begin{quote}
I want upbeat combat music for a 2D RPG set in a plains biome.
Add flutes, wind textures, and African drums to get the player ready
for battle. The intended gameplay scene is an upbeat 2D RPG battle
in an open plains biome.
\end{quote}

\begin{table}[htbp]
\centering
\small
\caption{A complete expert-revised rubric for RPG battle music,
comprising six base and six completion requirements.}
\label{tab:app-rubric-example}
\begin{tabular}{
    @{}
    p{0.12\linewidth}
    p{0.58\linewidth}
    p{0.24\linewidth}
    @{}
}
\toprule
Source & Requirement & Dimension \\
\midrule

Base
& upbeat combat music for a 2D role-playing game
& Style \\

Base
& open-plains biome atmosphere
& Mood \\

Base
& flute parts
& Instrumentation/vocal \\

Base
& wind textures
& Production/texture \\

Base
& African drum rhythms
& Rhythm \\

Base
& energizing battle-ready mood
& Mood \\

Completion
& open-air spaciousness
& Production/texture \\

Completion
& buoyant loopable battle structure
& Structure/energy \\

Completion
& call-and-response between flute and drums
& Rhythm \\

Completion
& steady forward motion without a long cinematic introduction
& Structure/energy \\

Completion
& bright and cheerful major-key tonality
& Harmony/melody \\

Completion
& drum patterns progressively intensifying to build battle readiness
& Rhythm \\

\bottomrule
\end{tabular}
\end{table}

% ============================================================
% B. Human assessment and evaluator agreement
% ============================================================

\section{Human Assessment and Evaluator Agreement}
\label{app:human-assessment}

Three music experts completed two human-assessment tasks: judging
individual rubric requirements and rating intent alignment across
generated candidates. For both tasks, assignments were divided
approximately equally among the experts without overlapping
assignments, and each expert independently assessed their assigned
samples. Experts received common task and scoring instructions before
annotation. They were instructed to listen to each clip in full before
rating and were allowed to replay it. System identities and automatic
scores were hidden in the annotation interface. Ratings concerned
fulfillment of the request rather than personal musical preference.

\paragraph{Rating anchors.}
The item-level task uses a 20-point scale, and the system-level task
uses a five-point scale. The anchors are summarized in
Table~\ref{tab:app-rating-anchors}. For the 20-point scale, experts
distinguish degrees of fulfillment within each range. The two
assessment tasks are scored separately using their respective scales.

\begin{table}[htbp]
\centering
\small
\caption{Scoring anchors for item-level and system-level
expert assessments.}
\label{tab:app-rating-anchors}
\begin{tabular}{@{}ccp{0.70\linewidth}@{}}
\toprule
20-point & 5-point & Interpretation \\
\midrule

1--4 & 1
& Not fulfilled: the required property is absent or clearly
contradicted. \\

5--8 & 2
& Slightly fulfilled: only weak evidence is audible, with most
of the requirement unmet. \\

9--12 & 3
& Partially fulfilled: recognizable evidence is present, but
substantial shortcomings remain. \\

13--16 & 4
& Mostly fulfilled: the main requirement is realized, with minor
shortcomings. \\

17--20 & 5
& Fully fulfilled: the requirement is clearly and sufficiently
realized, without an evident deviation. \\

\bottomrule
\end{tabular}
\end{table}

\subsection{Rubric-level annotation}

For item-level assessment, the interface presents an audio clip and
its rubric requirements. The instructions ask evaluators to judge
whether the audible content meets each requirement, rather than
whether they like the music. Experts use a 20-point scale for
item-level assessment. For an integer rating
$r\in\{1,\ldots,20\}$, the normalized score is
\[
h=(r-1)/19.
\]
Uncertain responses are excluded from numerical analysis.

The agreement study in Table~\ref{tab:evaluator-agreement} uses
1,536 clip--item ratings from 100 clips and 25 requests, with one
expert rating per target.

\subsection{Matching automatic scores to human targets}
\label{app:agreement-matching}

Each evaluator is matched to human targets by clip and rubric-item
identifiers. All evaluation methods in Table~\ref{tab:evaluator-agreement}
use the same 1,536 clip--item targets from 100 clips and 25 requests. Item-level Spearman correlation and
Kendall's $\tau_b$ use these matched pairs. Clip-level scores average
the matched rubric items within each clip for both the automatic and
human scores, then compute Pearson correlation, Spearman correlation,
and Kendall's $\tau_b$ across clips.

\paragraph{Pairwise accuracy.}
We compare unordered pairs of clips generated for the same request,
using their clip-level mean scores. Pairs tied under the human score
are excluded. A pair counts as correct only when the automatic score
orders the clips in the same direction as the human score; an
automatic tie on a human-nontied pair counts as incorrect.
Differences below $10^{-12}$ in magnitude are treated as ties.
The 25 requests provide 150 candidate pairs before excluding three
human ties, leaving 147 comparisons. These comparisons share clips
and should not be interpreted as 147 independent requests.

\subsection{System-level listening interface}

The system-level listening study covers all 100 MuRA-Bench
requests and 400 audio clips, with four clips per request: direct ACE-Step, ACE-Step with
MIRA ($B=9$), MiniMax music-3.0, and Suno v5.5. The listening task
presents the four candidates labeled A--D together with the request
and its intent criteria. Candidate order is stably shuffled for each
evaluator--request combination, so reloading a task preserves its
presentation order. The annotation interface uses anonymous candidate
labels and audio endpoints, without showing generation-system names
or automatic evaluator scores.

The listening form asks evaluators to rate explicit, inferred, and
overall intent on a five-point scale. The instructions focus on
fulfillment of the request rather than overall music quality.
The interface also allows tied best-candidate selections.
We aggregate ratings separately for overall, explicit, and inferred
intent to obtain the system-level results in
Figure~\ref{fig:human-evaluation}.

Explicit intent concerns requirements stated directly by the user.
Inferred intent concerns supplementary requirements derived from
references and context in the request. Overall intent is rated
independently as a holistic judgment of whether the music achieves
the user's intended outcome.

\paragraph{Aggregation and uncertainty.}
Each clip receives one expert rating for each intent dimension.
For each system and dimension, we compute the arithmetic mean
of the 100 ratings, giving each request equal weight.
Let $y_1,\ldots,y_N$ denote these ratings, with $N=100$.
The mean and sample standard deviation are
\[
\bar y=\frac{1}{N}\sum_{i=1}^{N}y_i,
\qquad
s=\sqrt{\frac{1}{N-1}\sum_{i=1}^{N}(y_i-\bar y)^2}.
\]
The error bars show two-sided 95\% Student-$t$ confidence intervals:
\[
\left[
\bar y-t_{0.975,99}\frac{s}{\sqrt{100}},
\quad
\bar y+t_{0.975,99}\frac{s}{\sqrt{100}}
\right].
\]
These intervals summarize variation across rated requests;
they do not measure inter-rater reliability.

% ============================================================
% C. Evaluation targets and score interpretation
% ============================================================

\section{Evaluation Targets and Score Interpretation}
\label{app:evaluation-details}

\subsection{Evaluator baseline implementations}

The agreement study compares four scoring procedures on individual
rubric requirements. Their inputs and outputs differ as follows;
matching and aggregation use the protocol in
Appendix~\ref{app:agreement-matching}.

\paragraph{CLAPScore.}
The baseline uses \texttt{laion/larger\_clap\_music\_and\_speech}
to measure audio--text embedding similarity for each clip and
rubric requirement. We use raw cosine similarity as the item score.

\paragraph{Prompting.}
MusicFlamingo receives the audio and one rubric requirement and
directly generates a rating from 1 to 5. The prompt instructs it to
rely on audible evidence, avoid unsupported inference, and return
\texttt{Score: <number>}. Its anchors range from minimal or no
audible support to clear and complete satisfaction.
The parsed rating $s$ is normalized as $(s-1)/4$.

\paragraph{Cascade.}
MusicFlamingo first generates one detailed caption per clip without
access to the target rubric requirement. The caption covers audible
style, instrumentation, vocals, rhythm, harmony, structure,
production, mood, and temporal development, while instructing the
model not to guess uncertain details. The same caption is reused
for all requirements associated with the clip. A text-only Kimi 2.6
judge then receives the caption and one requirement and rates their
match from 1 to 5. It is instructed to judge only explicit caption
evidence and not treat missing or unclear information as present.
The returned rating is normalized as $(s-1)/4$.

\paragraph{MF-AQA.}
MusicFlamingo receives the audio and a requirement-specific yes/no
question. Following the probability-based scoring formulation
described in the main text, its score is the normalized probability
of the yes alternative. Each requirement is evaluated directly
against the audio.

\subsection{Gold rubrics and online rubrics}

The expert-revised gold rubric $R^\star(x)$ defines offline benchmark
evaluation. MIRA separately constructs an online rubric $R$ from
the request and admitted tool evidence before search. Online
observations guide candidate refinement and selection, whereas
gold-rubric scores evaluate the selected audio. Gold items and human
grading labels are not supplied to the generation or refinement loop.

\subsection{Soft scores and repair thresholds}

MF-AQA evaluates each item as an audio question with yes/no
alternatives. The normalized probability is
\[
q_i(a)=
\frac{\exp(z_i^{\mathrm{yes}})}
{\exp(z_i^{\mathrm{yes}})+\exp(z_i^{\mathrm{no}})}.
\]
Online candidate selection averages these probabilities over the
fixed online rubric. The threshold $\eta=0.60$ identifies items
that may need repair; it does not binarize the probabilities used
for candidate ranking. Similarly, the memory tolerance
$\epsilon=0.03$ classifies changes between candidates as improvements,
regressions, or stable observations.

\subsection{Scope of the human comparisons}

The agreement study assesses requirement-level scoring and
within-request candidate ranking against expert judgments.
The listening study assesses overall, explicit, and inferred intent
directly from generated tracks. Both studies focus on intent
fulfillment; general perceptual quality and listener preference
are outside their scope.

% ============================================================
% D. Search and reproducibility
% ============================================================

\section{MIRA Search and Reproducibility Details}
\label{app:mira-implementation}

\subsection{Configuration and budget accounting}

The experiments use Kimi 2.6 as the text planner and MusicFlamingo
as the music verifier. The three refinement backends in
Table~\ref{tab:main-results} are ACE-Step v1.5 Turbo,
SongGeneration2 Large, and YuE2-3B, each evaluated with direct
prompting and MIRA at $B=3$ and $B=9$. For each of the three
backends, generation and verification run on a single NVIDIA A800.
The table additionally reports ten direct API baselines:
MiniMax music-2.6 and music-3.0, Suno v4, v4.5, v5, v5.5, and v6,
Mureka V9 and V9.5, and StepAudio 3 Music.
Table~\ref{tab:app-search-settings} summarizes the search settings.
The generation budget includes the initial candidate. It counts
generated candidates, not planner calls, retrieval calls, verifier
questions, or elapsed time.

\begin{table}[htbp]
\centering
\small
\caption{MIRA search settings and generation-budget conventions.}
\label{tab:app-search-settings}
\begin{tabular}{@{}p{0.43\linewidth}p{0.51\linewidth}@{}}
\toprule
Setting & Value or rule \\
\midrule

Generation budget $B$
& 3 or 9, including the initial candidate \\

Low-score feedback threshold $\eta$
& 0.60 \\

Memory change tolerance $\epsilon$
& 0.03 \\

Online ranking score
& Mean item satisfaction over the fixed online rubric \\

Complementary frontier criterion
& Mean of the lowest-scoring quarter of items \\

Final selection
& Highest-scoring successfully visited candidate \\

Memory scope
& Branch-specific; reset between requests \\

\bottomrule
\end{tabular}
\end{table}

\subsection{Search configuration and candidate selection}

Tool grounding and online rubric construction are performed before
candidate generation. The online rubric remains fixed throughout
the search for each request. With $B=9$, the controller allocates
one generation to the initial candidate, three to exploratory
candidates, up to four to expanding two selected parents, and one
to a final action from the highest-scoring candidate.
Each selected parent produces up to two children.
With $B=3$, the budget covers the initial candidate and two
exploratory candidates. Candidates in each expansion batch are
evaluated before the next batch is constructed.

Parent selection retains the candidate with the highest mean
requirement satisfaction and a distinct candidate with the highest
lower-tail satisfaction. The latter averages the lowest
$\max(1,\lfloor m/4\rfloor)$ item scores, where $m$ is the number
of online requirements. Final selection considers all evaluated
candidates, ranking them by mean satisfaction, followed by pass
rate, lower-tail satisfaction, minimum item score, and generation
order.

ACE-Step and YuE2 retain a fixed generation seed within each
request's adaptive search. SongGeneration2 uses its native
stochastic generation interface. For the final action, the
controller refines the strongest candidate; on SongGeneration2,
it may instead resample that candidate when the refinement
condition is not met. Vocal lyrics are supplied before execution
and remain fixed across candidates.

\subsection{Equal-budget and component configurations}

The equal-budget comparison and component analysis use the same
20 requests with ACE-Step v1.5 Turbo and YuE2-3B.
Direct generation produces one candidate from the original request.
Best-of-9 generates nine candidates from the unchanged original
request and ranks them using an online rubric decomposed from that
request. Tools + Best-of-9 performs tool grounding once, constructs
an online rubric from the augmented request, and holds both the
augmented request and rubric fixed across nine independent
generations. Sampling uses consecutive generation seeds for
ACE-Step and YuE2.

Tools + Adaptive search uses the same tool-grounding procedure and
budgeted search controller as full MIRA, with memory disabled.
Current candidate scores, satisfied and failed requirements, and
targeted verifier feedback remain available. Historical requirement
locks, successful repair phrases, persistent-failure records,
recent experience notes, and tree-contrast memory are omitted.
Full MIRA enables these memory inputs while retaining the same
search budget and controller.

All multi-candidate configurations use MusicFlamingo mean
satisfaction over their online rubric for final selection.
The expert-revised gold rubric is used only to evaluate the
selected output.
\subsection{Cross-evaluator Assessment}
\label{app:cross-evaluator}

We evaluate intent alignment with two additional evaluators,
MOSS-Music-8B-Instruct and Qwen3-Omni-30B-A3B.
The assessment covers direct generation and MIRA at
$B=3$ and $B=9$ on ACE-Step v1.5 Turbo,
SongGeneration2 Large, and YuE2-3B, using all 100
MuRA-Bench requests for every configuration.
MusicFlamingo guides search and candidate selection.
The alternative evaluators apply AQA scoring to the same
selected outputs against the expert-revised gold rubrics.
We additionally evaluate representative commercial
direct-prompting baselines with both evaluators.
Scores follow the aggregation protocol in the main text.

Table~\ref{tab:cross-evaluator} shows that both MIRA budgets
improve Overall, D-Macro, Base, and Comp. over direct
generation across all three backends under both evaluators.
At $B=9$, improvements are also observed in all seven
musical dimensions.
These results demonstrate consistent intent-alignment gains
under multiple evaluators.
Comparisons are made within each evaluator, since their
absolute score scales differ.

Evaluator agreement with human judgments provides additional
context for interpreting these results.
In Table~\ref{tab:evaluator-agreement}, Qwen3-Omni shows lower
agreement with expert ratings than MusicFlamingo across all
six metrics.
As a general-purpose multimodal model, Qwen3-Omni may be less
sensitive to some fine-grained musical requirements than a
music-specialized evaluator.
This is a possible explanation rather than a directly tested
cause of the agreement gap.
We therefore use its scores as complementary evidence,
while the stronger expert agreement of MusicFlamingo
supports its use as the primary evaluator.

\begin{table}[p]
\centering
\caption{Cross-evaluator assessment on MuRA-Bench.
MOSS-Music and Qwen3-Omni score outputs from direct generation
and MusicFlamingo-guided MIRA.
Commercial direct-prompting baselines are evaluated
with MOSS-Music and Qwen3-Omni.
Scores are multiplied by 100; higher is better.
Superscripts indicate differences from the reported Direct
score within each backend--evaluator group.
Bold marks the best reported score within each refinement
group or among the commercial baselines, before rounding.}
\label{tab:cross-evaluator}

\begingroup
\definecolor{miraBand}{RGB}{233,242,248}
\definecolor{miraHeader}{RGB}{211,226,237}
\definecolor{miraGain}{RGB}{55,125,70}
\definecolor{miraDrop}{RGB}{175,65,65}

\providecommand{\mup}[2]{}
\providecommand{\mdown}[2]{}
\renewcommand{\mup}[2]{%
  #1\textsuperscript{\textcolor{miraGain}{\ensuremath{\uparrow}#2}}}
\renewcommand{\mdown}[2]{%
  #1\textsuperscript{\textcolor{miraDrop}{\ensuremath{\downarrow}#2}}}

\setlength{\tabcolsep}{2.5pt}
\renewcommand{\arraystretch}{1.12}

\resizebox{\linewidth}{!}{%
\begin{tabular}{@{}l*{11}{c}@{}}
\toprule
& \multicolumn{4}{c}{\textbf{Intent alignment}}
& \multicolumn{7}{c}{\textbf{Musical dimensions}} \\
\cmidrule(lr){2-5}
\cmidrule(lr){6-12}
\textbf{Backend / Setting}
& Overall & D-Macro & Base & Comp.
& Style & I/V & Mood & Rhythm & H/M & S/E & P/T \\
\midrule

\rowcolor{miraHeader}
\multicolumn{12}{c}{\textbf{Evaluator: MOSS-Music}} \\

\rowcolor{miraBand}
\multicolumn{12}{c}{\textit{ACE-Step v1.5 Turbo}} \\

Direct
& 64.7 & 65.6 & 66.5 & 63.8
& 69.0 & 55.8 & 68.9 & 67.5 & 69.1 & 69.7 & 59.0 \\

MIRA ($B=3$)
& \mup{71.2}{6.5}
& \mup{71.5}{5.9}
& \mup{\textbf{73.1}}{6.7}
& \mup{68.1}{4.3}
& \mup{73.3}{4.3}
& \mup{\textbf{66.0}}{10.2}
& \mup{80.1}{11.2}
& \mup{70.5}{3.0}
& \mup{70.9}{1.8}
& \mup{70.6}{0.9}
& \mup{68.8}{9.8} \\

MIRA ($B=9$)
& \mup{\textbf{71.9}}{7.1}
& \mup{\textbf{72.7}}{7.1}
& \mup{73.1}{6.6}
& \mup{\textbf{69.8}}{6.0}
& \mup{\textbf{74.5}}{5.5}
& \mup{59.9}{4.1}
& \mup{\textbf{80.4}}{11.4}
& \mup{\textbf{72.2}}{4.7}
& \mup{\textbf{74.8}}{5.7}
& \mup{\textbf{73.8}}{4.2}
& \mup{\textbf{73.2}}{14.2} \\

\rowcolor{miraBand}
\multicolumn{12}{c}{\textit{SongGeneration2 Large}} \\

Direct
& 62.9 & 63.1 & 65.0 & 60.5
& 59.7 & 46.6 & 68.7 & 70.2 & 62.3 & 66.8 & 67.1 \\

MIRA ($B=3$)
& \mup{66.0}{3.1}
& \mup{66.4}{3.3}
& \mup{68.6}{3.6}
& \mup{62.2}{1.7}
& \mup{65.1}{5.4}
& \mup{47.1}{0.5}
& \mup{\textbf{78.4}}{9.6}
& \mdown{68.3}{1.9}
& \mup{67.8}{5.5}
& \mup{\textbf{69.8}}{3.0}
& \mup{68.2}{1.0} \\

MIRA ($B=9$)
& \mup{\textbf{69.8}}{6.9}
& \mup{\textbf{69.5}}{6.5}
& \mup{\textbf{71.8}}{6.8}
& \mup{\textbf{67.0}}{6.5}
& \mup{\textbf{74.2}}{14.5}
& \mup{\textbf{51.5}}{4.9}
& \mup{76.0}{7.3}
& \mup{\textbf{73.5}}{3.4}
& \mup{\textbf{68.7}}{6.3}
& \mup{69.3}{2.5}
& \mup{\textbf{73.7}}{6.5} \\

\rowcolor{miraBand}
\multicolumn{12}{c}{\textit{YuE2-3B}} \\

Direct
& 73.9 & 73.7 & 76.3 & 71.3
& 77.3 & 62.0 & 80.7 & 74.6 & 71.2 & 76.7 & 73.5 \\

MIRA ($B=3$)
& \mup{77.5}{3.6}
& \mup{77.7}{4.0}
& \mup{79.8}{3.4}
& \mup{76.0}{4.7}
& \mup{\textbf{82.6}}{5.4}
& \mup{69.8}{7.8}
& \mup{\textbf{84.6}}{3.9}
& \mup{77.4}{2.9}
& \mup{75.0}{3.8}
& \mup{79.7}{3.0}
& \mup{75.0}{1.5} \\

MIRA ($B=9$)
& \mup{\textbf{78.4}}{4.5}
& \mup{\textbf{78.1}}{4.4}
& \mup{\textbf{80.7}}{4.4}
& \mup{\textbf{76.8}}{5.5}
& \mup{79.8}{2.5}
& \mup{\textbf{71.4}}{9.4}
& \mup{84.0}{3.3}
& \mup{\textbf{78.4}}{3.8}
& \mup{\textbf{75.2}}{4.0}
& \mup{\textbf{81.3}}{4.5}
& \mup{\textbf{76.5}}{3.0} \\

\rowcolor{miraBand}
\multicolumn{12}{c}{
  \textit{Commercial systems (direct prompting)}
} \\

MiniMax music-3.0
& 78.4 & 78.3 & 80.4 & 77.5
& 80.1 & 67.5 & 81.4 & 78.3
& 79.9 & 79.4 & \textbf{81.4} \\

Mureka V9.5
& 76.5 & 77.2 & 78.8 & 74.9
& 80.4 & 65.6 & 82.2 & 78.3
& 77.7 & 80.1 & 76.3 \\

Suno v5
& 79.3 & 80.1 & 82.0 & 77.2
& \textbf{86.0} & 68.3 & 82.8 & 80.2
& 80.6 & 81.7 & 81.1 \\

Suno v5.5
& \textbf{80.0} & \textbf{80.8} & \textbf{82.4} & 77.7
& 85.7 & 69.7 & 82.4 & \textbf{83.0}
& 80.2 & 85.1 & 79.7 \\

Suno v6
& 79.7 & 80.7 & 82.2 & \textbf{77.9}
& 84.8 & \textbf{69.7} & \textbf{83.5} & 78.8
& \textbf{82.3} & \textbf{86.9} & 78.8 \\

\midrule
\rowcolor{miraHeader}
\multicolumn{12}{c}{\textbf{Evaluator: Qwen3-Omni}} \\

\rowcolor{miraBand}
\multicolumn{12}{c}{\textit{ACE-Step v1.5 Turbo}} \\

Direct
& 59.5 & 60.0 & 62.4 & 57.1
& 66.7 & 48.6 & 66.2 & 62.1 & 58.8 & 64.3 & 53.3 \\

MIRA ($B=3$)
& \mup{65.7}{6.2}
& \mup{67.4}{7.4}
& \mup{69.8}{7.4}
& \mup{62.9}{5.9}
& \mup{\textbf{74.2}}{7.5}
& \mup{53.3}{4.7}
& \mup{76.8}{10.5}
& \mup{66.0}{3.9}
& \mup{\textbf{65.0}}{6.2}
& \mup{72.7}{8.4}
& \mup{63.8}{10.5} \\

MIRA ($B=9$)
& \mup{\textbf{67.1}}{7.6}
& \mup{\textbf{68.2}}{8.2}
& \mup{\textbf{70.1}}{7.6}
& \mup{\textbf{64.7}}{7.6}
& \mup{74.0}{7.3}
& \mup{\textbf{53.3}}{4.7}
& \mup{\textbf{77.8}}{11.6}
& \mup{\textbf{67.4}}{5.4}
& \mup{64.4}{5.6}
& \mup{\textbf{73.0}}{8.7}
& \mup{\textbf{67.5}}{14.2} \\

\rowcolor{miraBand}
\multicolumn{12}{c}{\textit{SongGeneration2 Large}} \\

Direct
& 53.0 & 53.1 & 55.0 & 51.8
& 51.4 & 35.5 & 60.9 & 62.8 & 48.2 & 58.9 & 54.2 \\

MIRA ($B=3$)
& \mup{56.5}{3.6}
& \mup{57.2}{4.1}
& \mup{59.5}{4.5}
& \mup{53.8}{2.0}
& \mup{58.7}{7.3}
& \mup{39.5}{4.0}
& \mup{67.0}{6.2}
& \mup{63.5}{0.7}
& \mup{54.0}{5.7}
& \mup{61.1}{2.2}
& \mup{57.0}{2.8} \\

MIRA ($B=9$)
& \mup{\textbf{64.3}}{11.4}
& \mup{\textbf{64.4}}{11.3}
& \mup{\textbf{67.9}}{12.9}
& \mup{\textbf{61.6}}{9.8}
& \mup{\textbf{71.6}}{20.2}
& \mup{\textbf{48.9}}{13.5}
& \mup{\textbf{72.2}}{11.4}
& \mup{\textbf{69.6}}{6.9}
& \mup{\textbf{55.0}}{6.8}
& \mup{\textbf{66.4}}{7.5}
& \mup{\textbf{67.0}}{12.8} \\

\rowcolor{miraBand}
\multicolumn{12}{c}{\textit{YuE2-3B}} \\

Direct
& 66.8 & 68.2 & 70.7 & 62.7
& 74.3 & 57.9 & 77.3 & 65.5 & 64.1 & 71.8 & 66.2 \\

MIRA ($B=3$)
& \mup{69.0}{2.2}
& \mup{70.1}{1.9}
& \mup{71.5}{0.9}
& \mup{66.8}{4.1}
& \mup{75.0}{0.7}
& \mup{61.9}{4.0}
& \mup{79.7}{2.4}
& \mup{66.8}{1.3}
& \mup{\textbf{69.2}}{5.0}
& \mdown{71.3}{0.5}
& \mup{66.9}{0.7} \\

MIRA ($B=9$)
& \mup{\textbf{70.6}}{3.8}
& \mup{\textbf{71.5}}{3.3}
& \mup{\textbf{74.2}}{3.5}
& \mup{\textbf{67.5}}{4.8}
& \mup{\textbf{78.0}}{3.7}
& \mup{\textbf{62.3}}{4.4}
& \mup{\textbf{80.6}}{3.3}
& \mup{\textbf{69.0}}{3.5}
& \mup{68.1}{3.9}
& \mup{\textbf{73.7}}{2.0}
& \mup{\textbf{68.8}}{2.5} \\

\rowcolor{miraBand}
\multicolumn{12}{c}{\textit{Commercial systems (direct prompting)}} \\

MiniMax music-3.0
& 72.5 & 73.1 & 75.5 & 70.4
& 81.9 & 62.1 & 79.9 & 73.4 & 65.3 & 76.7 & 72.5 \\

Mureka V9.5
& 70.8 & 72.3 & 74.5 & 68.0
& 84.6 & 56.7 & 80.2 & 75.3 & 65.8 & 73.6 & 69.9 \\

Suno v5
& 73.5 & 74.6 & 77.2 & 70.6
& \textbf{87.1} & 61.5 & 81.3 & 74.7
& 68.6 & 76.3 & 72.6 \\

Suno v5.5
& 75.0 & 75.6 & 79.1 & \textbf{71.0}
& 87.0 & 63.8 & 82.9 & \textbf{77.3}
& 66.4 & 78.1 & \textbf{73.9} \\

Suno v6
& \textbf{75.3} & \textbf{76.2} & \textbf{80.0} & 71.0
& 85.8 & \textbf{66.3} & \textbf{83.6} & 73.2
& \textbf{70.1} & \textbf{80.8} & 73.8 \\
\bottomrule
\end{tabular}%
}

\par\vspace{3pt}
\raggedright\fontsize{7}{8.5}\selectfont
Both evaluators use AQA scoring.
D-Macro: dimension-macro; Base/Comp.: explicit/inferred intent.
I/V: instrumentation/vocal; H/M: harmony/melody;
S/E: structure/energy; P/T: production/texture.
\textcolor{miraGain}{$\uparrow$}: increase;
\textcolor{miraDrop}{$\downarrow$}: decrease relative to Direct.

\endgroup
\end{table}

\subsection{Trace fields for qualitative analysis}

The curated case records contain the original request, prompts
before and after revision, generation seeds, node/round identifiers,
search actions, verifier feedback, resolved and regressed
requirements, tool calls, and evidence cards. An evidence card
records its tool/query, claim, supporting text, confidence, and
any risk note. These fields allow a prompt edit to be inspected
alongside its stated motivation and observed rubric changes.

The trace links each branch action to its supporting evidence
and observed rubric changes.
Table~\ref{tab:component-ablation} evaluates component contributions,
while Figure~\ref{fig:bestof9-vs-mira} compares MIRA with independent
sampling under the same generation budget.

% ============================================================
% G. Prompt templates
% ============================================================

\clearpage
\section{Prompt Templates and Implementation Interfaces}
\label{app:prompt-templates}

The five templates below specify tool routing, evidence fusion,
online requirement decomposition, candidate refinement, and audio
verification in the reference implementation. Braced placeholders
denote runtime inputs, and JSON braces are shown as presented to
the model. Each template is supplied as one user-role message;
the verifier additionally receives audio. Memory fields are
populated only when the corresponding memory mode is enabled.

\subsection{Tools and evidence admission}

The router exposes three tools.
\texttt{music\_reference\_lookup} accepts an entity, reference type
(artist/song/album), and reason.
\texttt{scene\_to\_music\_lookup} accepts a scene query, optional
scene type, and reason.
\texttt{track\_audio\_evidence} accepts a query or local audio path
and a reason. The planner selects tools according to the
information required by each request.

The deterministic evidence gate requires nonempty claim and
support fields, a true \texttt{usable\_for\_prompt} flag, and
confidence at least 0.45. An addendum with blocked copy/imitate
wording or no accepted evidence card is rejected.
Reference-query strings are removed from the addendum before it
is appended to the original request. The gate operates on the
tool-added text; it does not remove references already present
in the user's request. Evidence admission uses field validation
and a confidence threshold; claim support is supplied by the
tool and fusion stages.

\subsection{Requirement and candidate interfaces}

Online requirement construction uses the structured decomposition
template below. Its online type tags include style, mood,
instrumentation, rhythm, structure, production, vocal, negative,
reference safety, and other. They are distinct from the benchmark's
seven gold-rubric dimensions.

Candidate refinement returns a complete replacement prompt.
The controller supplies up to eight lowest-scoring checks,
compacted to requirement text, probability, and status.
The prompt asks for fewer than 110 words; the output guard
additionally filters disallowed sentences, removes repeated
labels, and clips to 110 words. If fewer than eight words remain,
it constructs a fallback from the original request and failure
cues.

\subsection{Structured memory input}

Tree-structured memory is computed from recorded nodes and edges.
A routed view contains the selected round and strategy,
comparison rounds, relative aggregate score, sibling advantages
and weaknesses, parent--child improvements and regressions,
newly resolved or weakened items, and the
\texttt{preserve\_next}/\texttt{repair\_next} lists.
This JSON view fills \texttt{contrastive\_memory\_view} in the
candidate template when memory is enabled; otherwise the slot
is \texttt{None}.

\subsection{Model invocation and answer-token scoring}

The text-model interface sends the formatted text as one user-role
message. Kimi 2.6 uses a temperature of 0.6 with thinking disabled.

MF-AQA reads the next-token logits after the chat template's
generation prefix. It collects unique single-token encodings
of \texttt{yes}, \texttt{Yes}, \texttt{YES}, and their leading-space
variants, and constructs the analogous no-token set.
If these sets are $Y$ and $N$, the effective answer logits are
\[
z^{\mathrm{yes}}
=
\log\sum_{t\in Y}\exp(\ell_t),
\qquad
z^{\mathrm{no}}
=
\log\sum_{t\in N}\exp(\ell_t).
\]
Their normalized ratio gives the soft satisfaction probability
defined in the main text. Deduplicating token IDs avoids counting
the same answer token more than once.

\subsection{Core prompt templates}

The boxes reproduce the template instructions and output contracts.
Line wrapping is for presentation only; the accompanying text files
preserve the template content. Tool definitions and structured
observations are inserted into the indicated slots at runtime.

\begin{tcolorbox}[
    promptbox,
    title={Prompt \thesection.1 --- Tool routing},
    title after break={Prompt \thesection.1 --- Tool routing (continued)}
]
\begin{Verbatim}
You are an Anthropic-style tool router for a text-to-music prompt. Return ONLY JSON.

Tools: {tools_json}

Prompt: {prompt}

Context: {context_json}

Messages: {messages_json}

Policy:
- Named artist/song/album or style reference -> music_reference_lookup
- Specific local reference audio or exact track preview need -> track_audio_evidence
- Scene/function/visual/use-case intent -> scene_to_music_lookup only when that scene/function is vague and needs stylistic translation
- Do not call scene_to_music_lookup merely to restate explicit duration, instrumentation, breakdown, ending, BPM, or structure requirements
- Tool augmentation must preserve the original prompt unchanged and only expand vague reference/style/scene wording
- Use no tools for generic genre-only prompts with enough detail
- Max {budget} tool uses total

If tools are needed:
{"type":"tool_use","tool_uses":[{"id":"toolu_001","name":"music_reference_lookup","input":{"entity":"...","reference_type":"artist","reason":"..."}}]}

If no tools are needed:
{"type":"final","tool_uses":[],"evidence_cards":[],"prompt_addendum":"","notes":"No grounding needed."}

\end{Verbatim}
\end{tcolorbox}

\begin{tcolorbox}[
    promptbox,
    title={Prompt \thesection.2 --- Evidence fusion},
    title after break={Prompt \thesection.2 --- Evidence fusion (continued)}
]
\begin{Verbatim}
Fuse tool results into a compact text-to-music generation plan. Return ONLY JSON.

Prompt: {prompt}

Tool results: {tool_results_json}

Rules:
- Prefer tool_result.reference_plan over flat verified_traits.
- prompt_addendum: English, compact generation plan bullets, only expanding vague reference/style/scene parts.
- Do not restate or rewrite explicit user requirements from the prompt.
- Do not add duration, BPM, key, exact timestamps, bar counts, or new structure.
- Do not include artist names, song names, copy/imitate/exactly-like language
- If no usable evidence, prompt_addendum must be ""
- Every claim in prompt_addendum needs an evidence_card with matching support
- notes: short explanation

Shape:
{"type":"final","reference_plan":{"style_anchor":[],"arrangement_plan":[],"sound_design":[],"rhythm_groove":[],"mix_space":[],"mood_energy":[],"avoid":[]},"evidence_cards":[{"source_tool":"...","source_query":"...","claim":"...","support":"...","confidence":0.0,"usable_for_prompt":true,"risk":""}],"prompt_addendum":"","notes":""}

\end{Verbatim}
\end{tcolorbox}

\begin{tcolorbox}[
    promptbox,
    title={Prompt \thesection.3 --- Online requirement decomposition},
    title after break={
        Prompt \thesection.3 --- Online requirement decomposition (continued)
    }
]
\begin{Verbatim}
You are a music prompt analyst for a text-to-music generation loop.
Given a music generation prompt, extract concrete requirements as a JSON array of objects.

Each object MUST have exactly these keys:
- "text": short English requirement string
- "type": one of ["style","mood","instrumentation","rhythm","structure","production","vocal","negative","reference_safety","other"]
- "repair_priority": one of ["normal","low","violation_only"]

Rules:
- Use normal for positive requirements useful for prompt repair.
- Use low for subtle production/mix details or vague requirements.
- Use violation_only for negative constraints or reference safety.
- Keep each text concise and checkable by an audio caption model.
- Output JSON only.

Music prompt: {prompt}

\end{Verbatim}
\end{tcolorbox}

\begin{tcolorbox}[
    promptbox,
    title={Prompt \thesection.4 --- Standalone candidate refinement},
    title after break={
        Prompt \thesection.4 --- Standalone candidate refinement (continued)
    }
]
\begin{Verbatim}
You are the candidate generator in a text-to-music agent loop.
The critic stage has already converted audio-verifier scores into structured feedback.

Your job is NOT to append a small repair note. Create one complete alternate music prompt
that interprets the original request through the branch strategy while staying faithful.

Original request: {original_prompt}

Current music prompt: {current_prompt}

Current lyrics block: {current_lyrics}

Branch strategy: {branch_strategy}

Passed requirements: {passed}

Lowest failed/weak checks: {failed_checks}

Targeted MusicFlamingo QA: {targeted_qa}

Locked passed requirements: {locked_passed}

Locked successful repair phrases: {locked_repair_phrases}

Persistent failed requirements: {persistent_failed}

Recent iterative experience notes: {recent_experience_notes}

Tree-aligned parent-specific contrast memory: {contrastive_memory_view}

Rules:
1. The candidate prompt must be a standalone prompt that can replace the current music prompt.
2. Preserve the original request, vocal/instrumental intent, and locked passed requirements.
3. Use failed/weak checks and targeted QA only as priority cues; do not optimize only those cues.
4. If contrast memory is present, preserve its verified advantages, repair its routed weaknesses, and avoid
   intensifying edits associated with regressions. Treat numeric deltas as evidence, not text to copy.
5. Make this candidate audibly different from the current prompt according to the branch strategy.
6. Do not add new duration, timestamps, BPM, key, Hz/kHz, dB, bar counts, or exact mix numbers.
7. Do not introduce artist names or references that were not already in the original request.
8. Keep candidate_music_prompt English-only, concrete, musical, and under 110 words.
9. If current_lyrics is not "None", preserve it unless the failed requirements clearly need a small lyric change.
10. If current_lyrics is "None", return lyrics_block as an empty string.
11. Return ONLY JSON:
{"variant_focus":"...","critic_notes":["..."],"candidate_music_prompt":"...","lyrics_block":"","vocal_mode":"instrumental|vocal|auto"}

\end{Verbatim}
\end{tcolorbox}

\begin{tcolorbox}[
    promptbox,
    title={Prompt \thesection.5 --- MF-AQA verification},
    title after break={
        Prompt \thesection.5 --- MF-AQA verification (continued)
    }
]
\begin{Verbatim}
You are a strict music audio verifier. Listen to the audio and answer the question.
Use only audible evidence. If the requirement is ambiguous, weak, only implied, or not clearly audible, answer no.
Question: Does the audio clearly satisfy this requirement: "{requirement}"?
Answer with exactly one word: yes or no.
Answer:
\end{Verbatim}
\end{tcolorbox}
\section{Ethical Considerations}
\label{app:ethics}

\subsection{Data provenance and privacy}

MuRA-Bench is constructed from anonymized intent seeds collected
on the Mureka platform, with platform permission for research use
and public release of the curated requests. We removed usernames
and other personally identifying information, excluded requests
containing sensitive information, and retained content relevant
to musical intent. We release the curated benchmark rather than
raw user logs.

\subsection{Expert participation}

The participating music experts were informed that their revisions
and ratings would be used for research. Both rubric revision and
human evaluation followed shared task instructions, with experts
independently completing assigned subsets. During human evaluation,
system identities and automatic scores were hidden, and experts
were instructed to assess fulfillment of musical requirements
rather than personal preference.

\subsection{Musical references and creator rights}

Artist, song, and album references are interpreted as general
musical characteristics that can be assessed by listening, rather
than direct imitation targets. Their inclusion does not imply
endorsement or authorization from the referenced creators.
Intent-alignment scores assess fulfillment of musical requirements
and do not establish the originality or copyright status of
generated audio.

\subsection{Interpretive and cultural bias}

Implicit musical intent can admit multiple reasonable
interpretations. Our expert revision guidelines require contextual
support for completed requirements and the removal of personal
preferences and overly specific guesses. These checks constrain
unsupported interpretations, but model judgments and the
perspectives of three experts may still reflect particular musical
conventions. MuRA-Bench measures alignment with request-specific
rubrics, rather than a universal standard of musical quality or
artistic value.

\end{document}